# Tailoring of Charge Carriers with Deposition Temperature in Pulsed Laser Deposited BiFeO$_3$ thin films

*R S Viswajit[a,b], K Ashok[b], K B Jinesh[a*]*

[a] Electronic Materials and Devices (EMERALD) Laboratory, Department of Physics, Indian Institute of Space Science and Technology (IIST), Thiruvananthapuram, Kerala 695547, India

[b] Vikram Sarabhai Space Centre (VSSC), Indian Space Research Organisation (ISRO), Thiruvananthapuram, Kerala 695022, India

## Abstract

This research investigates the structural, topological, electrical and optical properties of pulsed laser deposited polycrystalline BiFeO$_3$ thin films on silicon and glass substrate at varying deposition temperatures ranging from 400°C to 700°C. X-ray diffraction confirm rhombohedral phase and X-ray photoelectron spectroscopy reveals stoichiometric BiFeO$_3$ films. The optical bandgap of thin films obtained from absorption spectra increases with the substrate temperature. Photoluminescence emission spectrum reveals the defects and Atomic Force Microscopy analysis bring out the surface topography and crystallinity improvement with temperature. The charge transport studies reveal a transition in conductivity from n-type at lower deposition temperature to p-type at higher deposition temperature, attributed to oxygen and bismuth vacancies respectively. The intricate understanding of conductivity tuning and the leaky nature of BiFeO$_3$ thin film opens avenues for applications in non-volatile memories, particularly neuromorphic devices.

*Corresponding author email: kbjinesh@iist.ac.in

**Keywords:** Pulsed Laser Deposition, BiFeO$_3$, Oxygen Vacancy, Trap Density, Schottky Barrier.

# 1. Introduction

Bismuth ferrite - BiFeO$_3$ or BFO is an extensively studied single phase multiferroic perovskite material that retain its bulk properties even at the thin film level. Multiferroics simultaneously exhibit more than one ferroic order through spontaneous ordering such as ferromagnetism, ferroelectricity and ferroelasticity [1-3]. BiFeO$_3$ carries ferroelectric and antiferromagnetic properties with a strong magnetoelectric coupling that has potential applications in multistate random access memories [4-7] and spintronic devices [8,9]. The ferroelectric phase transition in BiFeO$_3$ occurs at its curie temperature (T$_C$) of 830°C, due to the large ionic radius of *Bi* and its *6s2* lone pairs, while the G-type antiferromagnetic phase transition occurs at its Neel temperature (T$_N$) of 370°C, attributed to the magnetic ordering of the *Fe-3d* electrons, which makes BiFeO$_3$ an ideal material for high temperature applications [10,11]. The magnetoelectric effect of BiFeO$_3$ enables the switching of magnetization with the electric field and makes it a promising material for binary information storage [3,12,13]. It also exhibits a high switchable ferroelectric polarization of 90-100 C/cm$^2$ along the [111] direction, which makes it suitable for non-volatile memory devices such as FeRAMs [6,7,11], logic gates [14,15] and ferroelectric capacitors [16-18]. The multifunctional properties of this material offer numerous opportunities for the development of next generation field-controlled devices. However, researchers are actively working to overcome the challenges associated with the leakage current and high density of defects in bismuth ferrite materials [19-21]. The physical and electrical properties of perovskites are affected by the presence of various defects in the material. One of the most common point defects in BiFeO$_3$ thin films are the vacancy defects like oxygen vacancies (V$_O$), bismuth vacancies (V$_{Bi}$) and iron vacancies (V$_{Fe}$). The defects with lower formation energy prevailing under the specific growth conditions gain dominance [22-24].

In this work, we explore the influence of vacancy defects on the functional properties of bismuth ferrite thin films. The $BiFeO_3$ films are synthesized using the pulsed laser deposition (PLD) technique, a highly versatile method allowing flexibility in target materials, stoichiometry, growth rate and the control over elemental vacancies [25-27]. In PLD, a pulsed laser beam ablates the target, the ablated material further interacts with the laser by direct photoexcitation and inverse Bremsstrahlung scattering. This interaction generates high energy plasma, which expands adiabatically in a reactive or non-reactive background gas, ultimately depositing onto the substrate. The key parameters controlling the PLD process include the laser wavelength, fluence, background gas concentration, vacuum pressure and deposition temperature ($T_D$). The deposition or substrate temperature is a critical parameter that controls the substrate's surface free energy, enabling the incoming atoms to bond better with the substrate and improving crystallinity and grain size. The substrate remains hydrophobic at low temperatures, causing weaker bonds with incoming species, resulting in smaller grain sizes with defects. In contrast, volatile species escape from the film at very high temperatures, resulting in vacancies [9,21,28]. Eventually, the deposition temperature controls the film's overall physical and electrical properties. The Schottky barrier formed by the contact electrodes also significantly regulates the transport characteristics of charge carriers [29,30]. In $BiFeO_3$ devices, the migration of $V_O$ leads to fluctuations in this energy barrier, resulting in interface-type multilevel neuromorphic memories that have excellent reproducibility and device uniformity [31,32]. Amidst the growing importance of brain-inspired electronics in AI and machine learning, this research unveils a significant advancement by precisely modulating the carrier types and defects through the PLD deposition temperature. This marks a paradigm shift in the fabrication approach of advanced electronic materials for neuromorphic devices.

## 2. Experimental
### 2.1. Preparation of BiFeO$_3$ thin films

The BiFeO$_3$ thin films are synthesized via the PLD technique onto glass and silicon substrates using a *99.9%* pure BiFeO$_3$ target sourced from ACI Alloys, Inc. (San Jose, USA). The substrate samples are cleaned using the three-cycle method involving 30-minute sonication in acetone, isopropyl alcohol and deionized water. The effects of deposition temperature on the material and electrical properties are investigated by depositing BiFeO$_3$ at four different temperatures, 400, 500, 600 and 700°C and samples are designated with deposition temperatures like BFO_400, BFO_500, etc. P-type silicon and glass substrates are used for material analysis and the $n^{++}$ silicon substrate is used for electrical characterization. PLD system uses a Q-switched solid state Nd:YAG laser (Quanta Ray, Spectra-Physics) operating at its third harmonics of *λ = 355 nm* with a pulse width of *8 ns* and repetition rate of *10 Hz*. The laser system is configured for a beam power of *0.4 W*, a spot size of *1 mm²* and the target-substrate gap kept at *4 cm*. The deposition process is carried out in a pure oxygen environment at *0.3 mbar*, with an initial base pressure of *10$^{-6}$ mbar* with a target thickness of *100 nm*. The deposited film is subjected to an additional annealing process at the respective deposition temperature. The film thickness is measured using Bruker DektakXT stylus profiler and the found thickness is *100±5 nm*. Thermally evaporated gold is deposited on BiFeO$_3$ film using a circular dot mask to form the top electrode of the devices.

### 2.2. Characterizations of BiFeO$_3$ thin films

Crystal structure analysis of the deposited films was carried out using X-ray diffraction (XRD, Bruker D8 ADVANCE diffractometer) with Cu-Kα excitation at *λ = 1.5406 Å* in the range of *20-60°* with *0.01°* step size and scan duration of 150 mins per sample. X-ray photoelectron spectroscopy (XPS) analysis using Thermo Scientific ESCALAB with polychromatic Al Kα radiation was used to study the chemical states of Bi, Fe, and O in the BiFeO$_3$ films. Atomic Force Microscopy (AFM) analyses were carried out using NaioAFM,

Nanosurf-AFM to study the surface morphology of the films. Photoluminescence (PL) spectra of the samples were recorded with the Horiba Fluromax 4 spectrophotometer at an excitation of *λ = 400nm* to find the defect levels in the films. UV-VIS absorption spectroscopy measurements to study the absorption characteristics and bandgap properties were done using LABINDIA UV 3200 UV-VIS spectrophotometer. The charge transport properties were examined by current-voltage (I-V) characterization approach using the Agilent B1500a parametric analyser.

## 3. Results and Discussion

### 3.1. Structural, Chemical and Morphology Characterizations

The crystallinity of $BiFeO_3$ films is studied using the X-ray diffraction technique. XRD patterns of the thin films deposited at 400 °C, 500 °C, 600 °C and 700 °C, alongside reference peaks, are shown in Fig. 1a. All the samples exhibit a polycrystalline structure of $BiFeO_3$, with sharp peaks indexed as *[104], [110], [024], [116], [112], [018]* and *[214]* corresponding to the rhombohedral phase with R3c space group (*JCPDS 72-2035*) [33]. In Fig. 1b, the close-up view reveals the characteristic split peak of $BiFeO_3$ at 32° corresponding to *[104]* and *[110]* crystal planes, along with shifted peak positions. Defects such as elemental vacancies in thin films introduce lattice strain, distorting the rhombohedral phase of $BiFeO_3$ and causing shifts in the XRD peaks. A left shift in the peak indicates tensile stress, while a right shift indicates compressive stress in the lattice. Notably, the films deposited at lower temperatures exhibit a slight right shift in the *2θ* peaks (Fig. 1b). The crystal size and lattice strain are calculated using Full Width at Half Maximum (FWHM) of XRD peaks and plotted against deposition temperature, aiming to understand their dynamics with change in the deposition temperature. In Fig. S1b, the lattice strain decreases with the increase in deposition temperature, primarily due to the reduction in defect densities. According to Kossar et al. [34], oxygen vacancy based defects induce compressive stress on the $BiFeO_3$ crystal lattice, leading to a right shift in the XRD peaks. Fig. S1a shows that the crystallite size increases with the deposition temperature, emphasising crystallinity improvement. X-ray photoelectron spectroscopy identifies the chemical state of individual atomic species in the 400°C and 700°C BiFeO$_3$ films. XPS survey scan is done at *1 eV* resolution and the individual element narrow scan is obtained at a higher resolution of *0.1 eV*. The binding energy scale of the spectrum is calibrated with the *C-1s* peak at *284.6 eV* [35]. Shirley's model was used for background compensation, which arises from the inelastic scattering of the photoelectrons [36]. XPS survey scan of both the films are shown in Fig. S1 and the corresponding narrow scan spectra of Bismuth, Iron and Oxygen are shown in Fig. 2 [35].

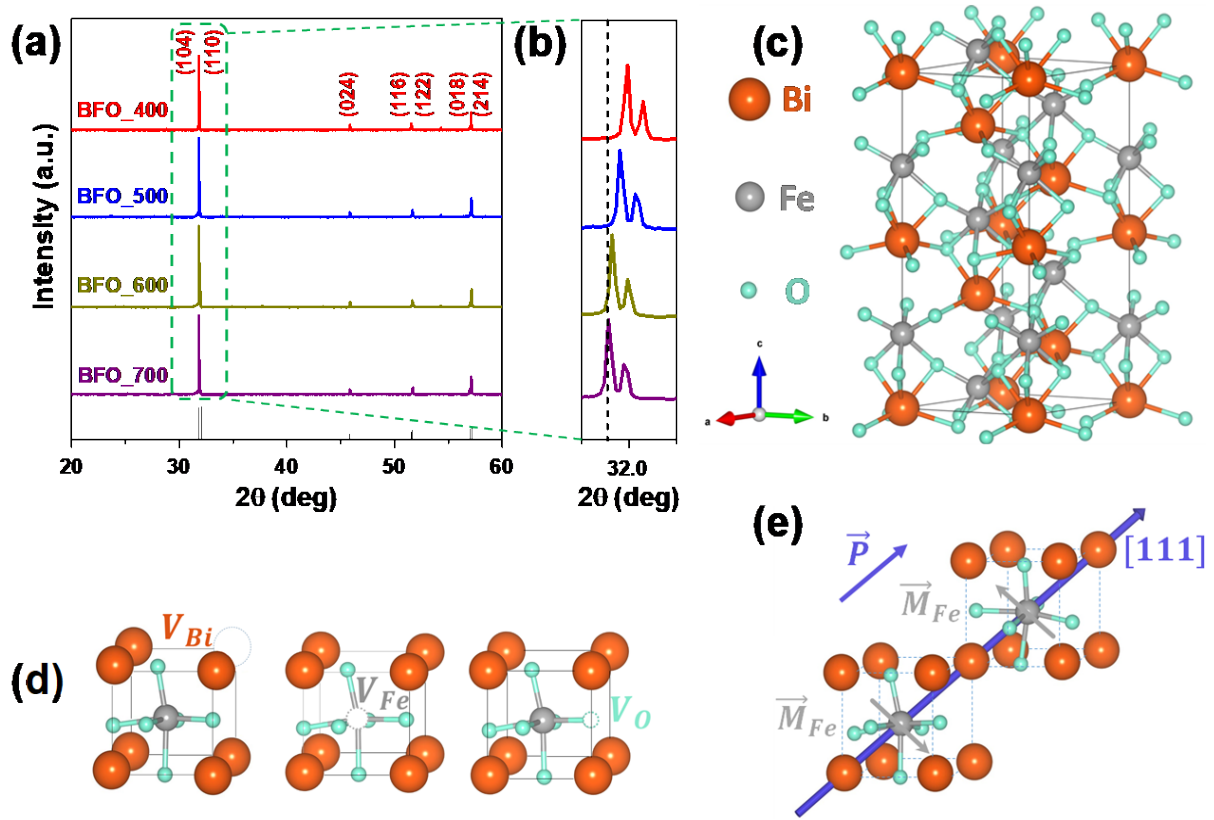

Fig.1: (a) XRD pattern of BiFeO$_3$ thin films on Si substrate with deposition temperature of 400 °C, 500 °C, 600 °C and 700 °C (b) The shift in peaks due to lattice strain induced by defects (c): The rhombohedral crystal structure of BiFeO$_3$, (d): the individual elemental vacancies $V_{Bi}$, $V_{Fe}$ and $V_O$ in the crystal unit cell of BiFeO$_3$ and (e): blue arrow along *[111]* indicates the direction of spontaneous polarisation, whereas the grey arrows indicates the magnetic moment for antiferromagnetism

The Bi-4f narrow scans (Fig. 2a, d) reveal the presence of $Bi^{3+}$ with ABO$_3$ structure, as evidenced by the binding energy peaks and spin-orbit splitting (5.33 eV) of *Bi-4f$_{7/2}$* and *Bi-4f$_{5/2}$* [35]. The *Fe-2p* narrow scan (Fig. 2b, e) confirms $Fe^{3+}$ predominance over $Fe^{2+}$ in both the samples as indicated by the deconvoluted peaks of *Fe-2p$_{3/2}$* and *Fe-2p$_{1/2}$*, spin-orbit splitting (13.8 eV) and their respective satellite peaks [35]. The satellite peaks (shake-off and shake-up peaks) result from a sudden shift in the lattice Columbic potential, which is caused by the influence of the ejected photoelectrons on the valence band. The satellite peaks corresponding to those of $Fe^{2+}$ are also termed surface peaks, which occur due to the crystal field loss in the film [35]. The presence of $Fe^{2+}$ indirectly indicates the presence of V$_O$, and the strength of their peaks reflects the density of V$_O$ [35]. The reduced intensity of satellite and surface peaks of *Fe-2p* in BFO_700 compared to BFO_400 points to a decrease in the density of oxygen vacancies at higher deposition temperature. Surface peaks are broader and less intense than the main peaks, as they arise from a smaller population of atoms or molecules.

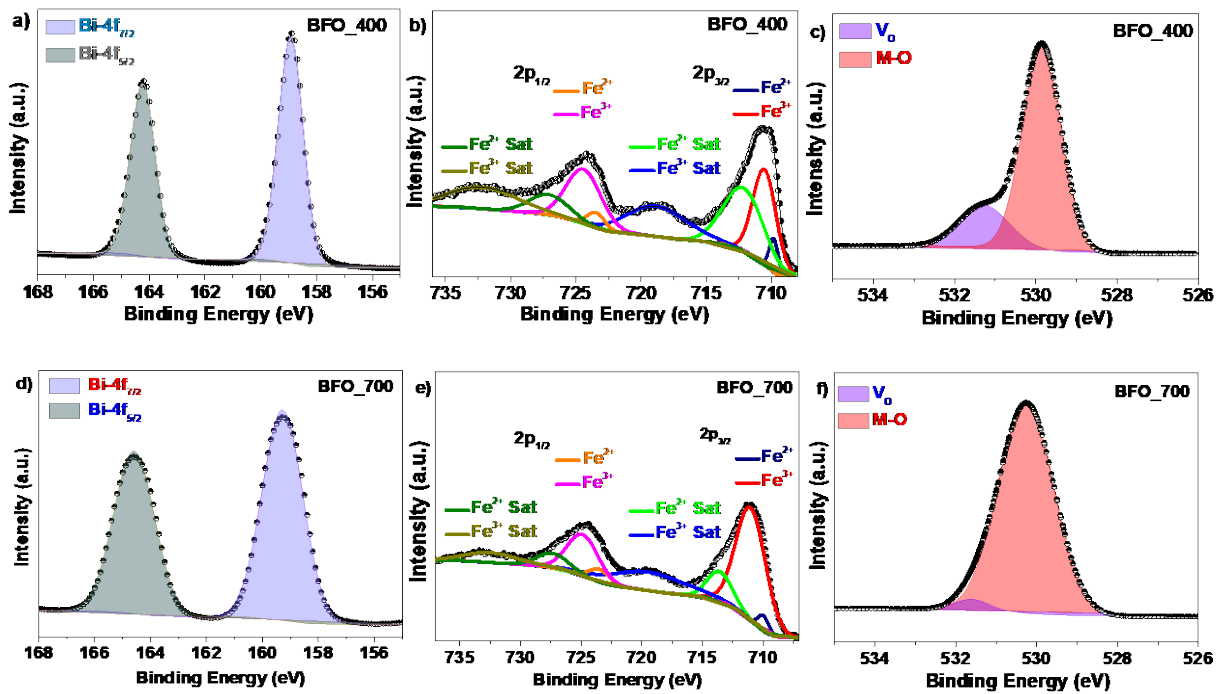

**Fig. 2**: XPS narrow scan of BiFeO$_3$ films on Si substrate, (a, d): *Bi-4f* spectrum showing spin-orbit split peaks, (b, e): *Fe-2p* spectrum and (c, f): *O-1s* spectrum, showing oxygen vacancies V$_O$ and metal connected oxygen M-O of BFO_400 and BFO_700

|  | Binding Energy (eV) |  | Composition |
| --- | --- | --- | --- |
|  | BFO_400 | BFO_700 |  |
| **Bi-4f$_{7/2}$** | 158.90 | 159.26 | Bi$^{3+}$ |
| **Bi-4f$_{5/2}$** | 164.23 | 164.59 |  |
| **Fe-2p$_{3/2}$** | 709.89 | 709.98 | Fe$^{2+}$ |
|  | 710.62 | 710.96 | Fe$^{3+}$ |
|  | 712.38 | 713.56 | Surface Peaks of 2p$_{3/2}$ |
|  | 719.2 | 719.45 | Satellite Peaks of 2p$_{3/2}$ |
| **Fe-2p$_{1/2}$** | 723.59 | 723.50 | Fe$^{2+}$ |
|  | 724.53 | 724.85 | Fe$^{3+}$ |
|  | 727.26 | 727.33 | Surface Peaks of 2p$_{1/2}$ |
|  | 732.7 | 732.45 | Satellite Peaks of 2p$_{1/2}$ |
| **O-1s** | 529.84 | 530.25 | Metal Oxygen |
|  | 531.29 | 531.60 | Oxygen Vacancies |

**Table 1**: Binding Energies and composition of elements in BiFeO$_3$ thin film, peak positions and splitting matches with Gomez et al. [35]

|  | BFO_400 | | | BFO_700 | | |
| --- | --- | --- | --- | --- | --- | --- |
|  | Composition (%) | Atomic Ratio | Chemical Formula | Composition (%) | Atomic Ratio | Chemical Formula |
| **Bismuth** | 17.64 | 0.97 |  | 18.38 | 0.93 |  |
| **Iron** | 18.14 | 1 | Bi$_{0.97}$FeO$_{2.73}$ | 19.72 | 1 | Bi$_{0.93}$FeO$_{2.99}$ |
| **Metal Oxygen** | 49.59 | 2.73 |  | 59.09 | 2.99 |  |
| **Oxygen Vacancy** | 14.58 | - |  | 2.81 | - |  |

**Table 2**: Percentage composition, Atomic ratio and Chemical formula for BFO_400 and BFO_700 films

In both the samples, the *O-1s* narrow scan (Fig. 2c, f) reveals the presence metal connected oxygen at around *530 eV* and oxygen vacancies at around *531 eV*, in agreement with the previous studies [35]. A significant reduction in the intensity of the oxygen vacancies is observed in the 700°C film compared to the 400°C film. This is attributed to the improved nucleation, stoichiometry and crystal formation process at higher deposition temperatures. Table 1 presents the chemical composition and spectrum peak identification of BiFeO$_3$ films, while Table 2 illustrates the percentage composition and atomic ratios. The oxygen vacancies dominate in BFO_400, whereas bismuth vacancies are introduced in BFO_700. This indicates a substantial improvement in stoichiometry and a decrease in defects, particularly those associated with oxygen vacancies at higher deposition temperatures, aligning with the XRD results. AFM mapped surface topography of the BiFeO$_3$ thin films is shown in Fig. 3, illustrating the corresponding average and RMS roughness. With the increase in deposition temperature, the higher surface energy activates the mobility of nanoparticles, resulting in conglomerated large grain structures [37]. The linear increase in the average and RMS roughness of the deposited film indicates improvement in crystallinity with increasing deposition temperature. AFM results of BFO_500 and BFO_600 samples are given in Fig. S2.

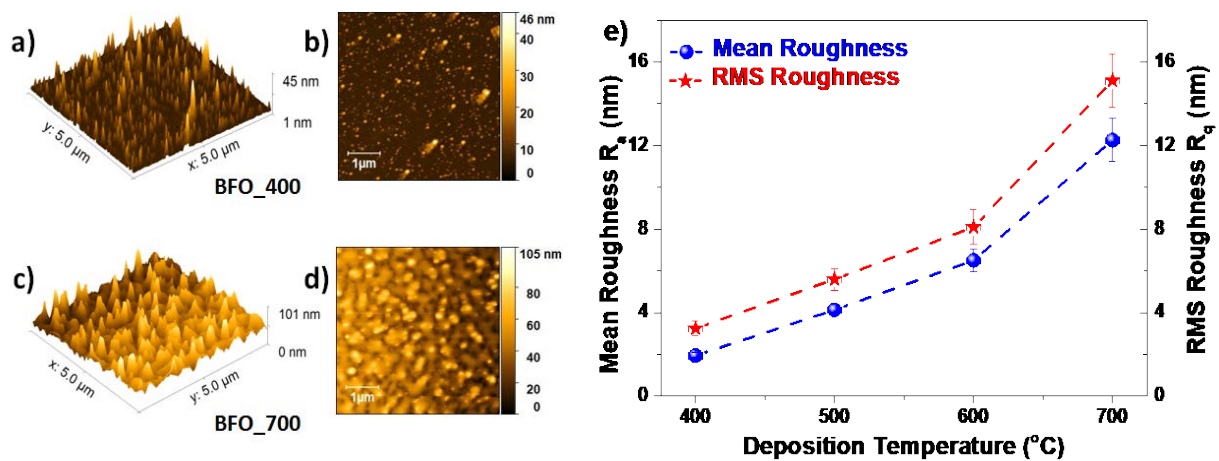

**Fig. 3**: (a, c): Dynamic mode AFM 3-D surface image, (b, d): 2-D images profile of BFO_400 and BFO_700 thin films on Si substrate with Scan area: 5×5 μm$^2$ and (e): the variation of the mean roughness (R$_a$) and RMS roughness (R$_q$) of the film surface with deposition temperature

### 3.2. Optical Emission and Absorption Characterizations

Photoluminescence (PL) spectra of the BiFeO$_3$ thin films identify peaks associated with defects such as cation vacancies, anion vacancies and interface defects. In addition to band-to-band transitions, the observed spectrum also reveals radiative recombinations from near band-edge and defect-level transitions, as shown in Fig. 4 and Fig. S3. The individual spectrum is deconvoluted to identify distinct peaks related to the transition between band edges and various defect levels. The energy levels of the vacancy defects are derived from the first-principle calculations of Shimada et al. [38] and Clark et al. [39] to qualitatively identify the respective radiative transitions. Oxygen vacancies exist in two separate states near the conduction band; $V_O^+$ (+/0) as a shallow trap and $V_O^{2+}$ (2+/+) as a deep trap. Bismuth vacancy primarily exists in $V_{Bi}^{3-}$ (0/3-) state as a shallow trap close to the valence band, while iron vacancy is in $V_{Fe}^{3-}$ (0/3-) state as a deep trap closer to the conduction band. Table S1 summarizes the dominant emission peaks from PL spectra and respective radiative recombinations between band edges and various defect levels are identified, indicating the presence of vacancy defects in all the BiFeO$_3$ films.

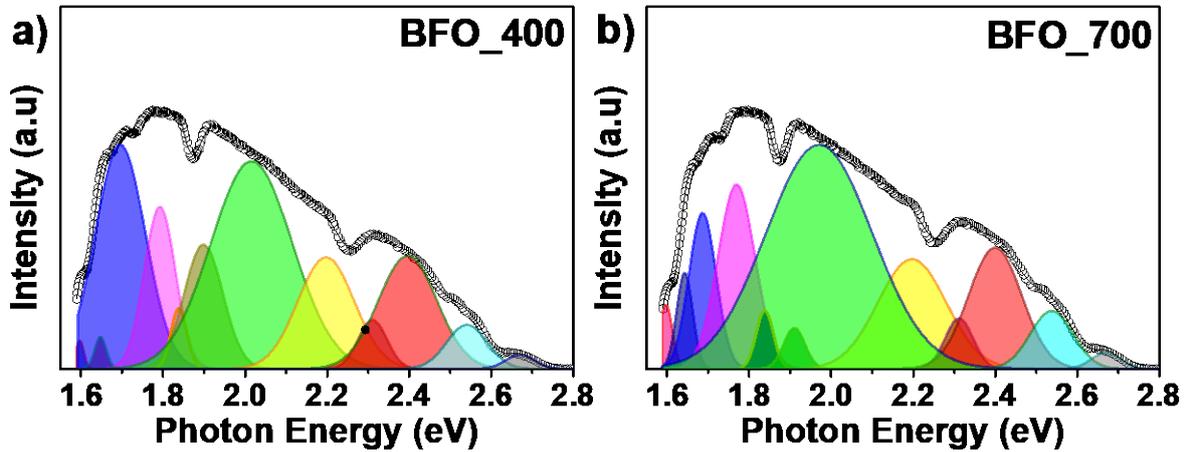

**Fig. 4**: Photoluminescence spectra of BiFeO$_3$ thin films on glass Substrate, (a): BFO_400 and (b): BFO_700 films at *400 nm* excitation

The bandgaps of the BiFeO$_3$ samples are estimated using UV-Visible absorption spectroscopy. Spectrum analysis indicates a direct optical bandgap (E$_g$) for the films as estimated from tauc plots (Fig. S4), exhibiting a discernible correlation with the deposition temperature (Fig. 5). The obtained $E_g$ values closely align with previously reported theoretical and experimental findings [40,41]. In BiFeO$_3$ films, the oxygen vacancies lead to defect levels within the bandgap, which are energetically close to the conduction band [42]. A higher concentration of oxygen vacancies leads to the overlapping of V$_O$ defect levels with the conduction band, effectively reducing the bandgap of the film [43,44].

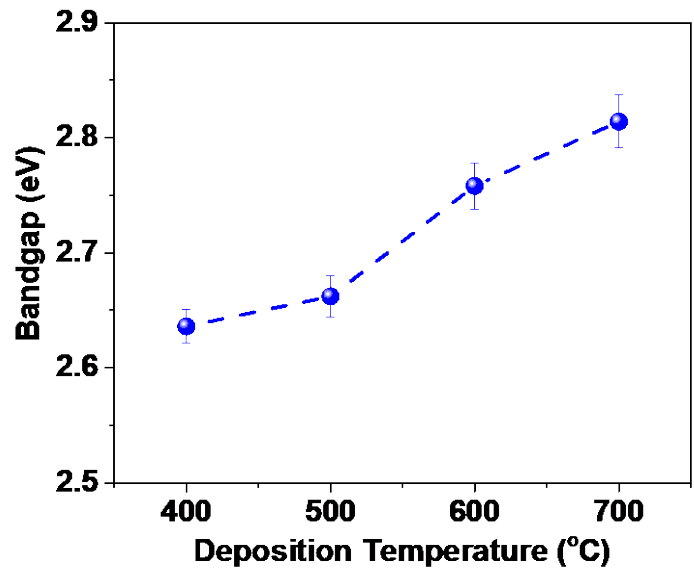

**Fig. 5**: Direct optical bandgap of BiFeO$_3$ thin films for different deposition temperatures, estimated from the absorption spectra of the thin films on the glass substrate

### 3.3. Electrical Characterization

The electrical properties of BiFeO$_3$ thin films are investigated through current-voltage (I-V) characterization of metal-insulator-metal devices in Au/BiFeO$_3$/n$^{++}$Si configuration. The dielectric does not allow charge transport due to its high bandgap. However, the presence of defects and traps in oxide based insulators can result in a leakage current [19,20]. The drift of free charge carriers, trapping and de-trapping of carriers at the defect sites (Space charge limited current or SCLC), thermionic emission, trap-assisted field-induced charge injection (Poole-Frenkel charge injection), electric field-assisted (Fowler-Nordheim) tunnelling and direct quantum tunnelling are the few mechanisms responsible for charge transport in oxides moderated by electric field and temperature [19,20,45]. At metal-oxide junctions, the Schottky barrier or conduction band (CB) offset hinders the free flow of charge carriers across the junction, serving as an energy barrier. Charge carriers overcome this energy barrier through hopping and the probability of conduction depends on the height of the barrier. This mode of conduction is referred to as Schottky thermionic emission, expressed by Eq. (1) [19,45,48].

$$J_S = AT^2 exp\left(\frac{-\varphi_b + q\sqrt{qE/4\pi\varepsilon_r\varepsilon_o}}{kT}\right) \qquad (1)$$

where T is Temperature, k is Boltzmann constant, E is the electric field, $\varepsilon_r$ is the dynamic permittivity and $\varphi_b$ is the conduction band offset.

The CB offset of the Au-BiFeO$_3$ junction and the dynamic permittivity ($\varepsilon_r$) of the oxide film is extracted from the offset and slope of the linear region of $ln(J/T^2)$ $vs$ $\sqrt{E}$ plot (Fig. S5) and related calculations are detailed in S5 (supplementary). The computed CB offset values of thin films (Fig. 6a) range between *0.7 eV* to *0.95 eV*. These values align with the theoretical results of Clark et al. for the Au-BiFeO$_3$ interface [49] and are consistent with the experimental findings of Chen et al. [50]. The increase in the barrier height observed with deposition temperature suggests a corresponding elevation in the conduction band edge of BiFeO$_3$. This observation matches with the concurrent rise in the optical bandgap, as confirmed by the results obtained from the UV-VIS absorption spectroscopy (Fig. 5). The computed dynamic permittivity $\varepsilon_r$ for the respective films are tabulated in Table S2, which are further used in calculating trap densities. In the conduction model, at very low voltage, the free charge carriers in oxides alone participate in conduction, following the relation $I \propto V^n$, where *n ≈ 1* is the Ohmic conduction. At slightly higher voltages, charge carriers undergo trapping and de-trapping at defect sites, resulting in an average current, termed as Space Charge Limited Current (SCLC), where *n ≈ 2* (Child's law). As a result of the trapping of all defect sites, the

SCLC current saturates at a critical voltage known as the Trap Filled Limited Voltage ($V_{TFL}$) (calculations in S6). The density of traps ($N_T$) is directly related to $V_{TFL}$ and the dynamic permittivity ($\varepsilon_r$) of the oxide [46], as described in Eq. (2).

$$N_T = \frac{2\varepsilon_o \varepsilon_r V_{TFL}}{qd^2} \quad (2)$$

where d is the thickness of the oxide film, $\varepsilon_r$ from Table S2 and $V_{TFL}$ from Fig. S6.

The PLD grown oxide films are susceptible to vacancy defects, particularly oxygen vacancies. The material characterization has demonstrated that at lower deposition temperatures (400°C and 500°C), the BiFeO$_3$ films exhibit lower crystallinity and non-stoichiometry and carry a large density of oxygen vacancies [51]. With increasing deposition temperature, the density of $V_O$ decreases due to improved nucleation and crystallization conditions. At 700°C, bismuth vacancies come into play, as their defect formation energy is lower than that of oxygen vacancies [47], which contributes to the total trap density in the BiFeO$_3$ thin films. The estimated trap densities from Eq. (2) reduce with increasing deposition temperature, as shown in Fig. 6b, affecting the conductivity of the films. These trap density values are comparable with previous experimental findings for bismuth ferrite thin films [46], which is also consistent with the XRD results in 3.1.

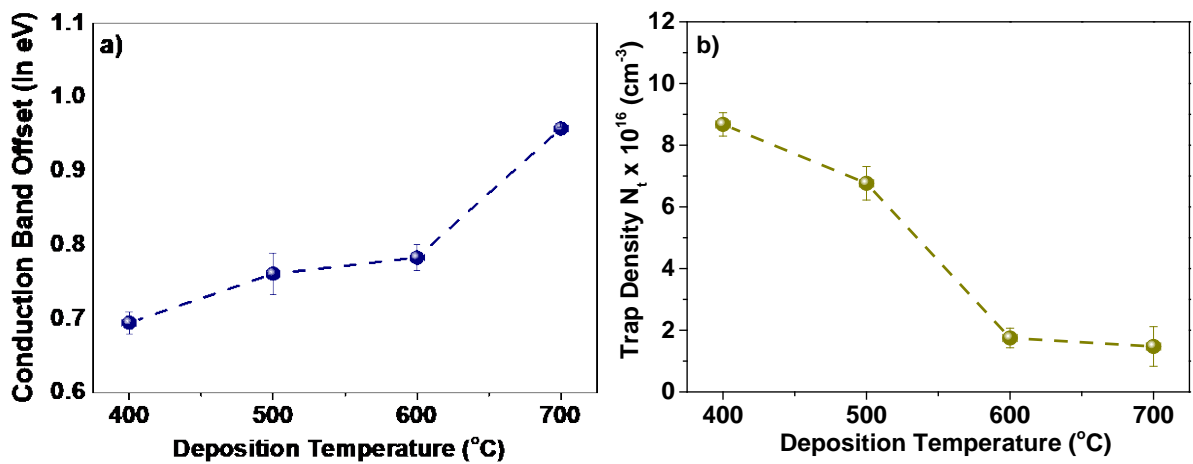

**Fig. 6:** (a): Conduction Band offset variation with deposition temperature, (b): Density of defects as a function of deposition temperature of the BiFeO$_3$ thin films

The nature of defects (acceptor or donor) and their activation energies help to understand the underlying conduction phenomenon in the BiFeO$_3$ thin films [20,22,23,52]. The effect of temperature on the activation of traps or defects is studied using the Arrhenius method. This method gives the nature of the traps and their activation energies (ΔE), which eventually contribute to the leakage current [5,53]. Eq. (3) expresses the current density (J) as a function of ΔE and device temperature (T) [45].

$$J = J_o exp\left(\frac{-\Delta E}{kT}\right) \quad (3)$$

The I-V measurements on devices at different temperature is used to generate Arrhenius plots $ln(J)$ $vs$ $1000/T$ (Fig. S7). The slope of the individual Arrhenius curve gives the activation energy of the associated traps at a particular bias voltage. Applying an electric field across the device results in the accumulation of opposite charges across the oxide layer. The interplay between the Coulombic force and the applied electric field across the oxide layer leads to the bending of energy bands, ultimately reducing the Schottky barrier at the metal-oxide junction. The band bending becomes more pronounced with the increasing bias, leading to a reduction in the activation energy of donor-type traps and an increase in the activation energy of acceptor-type traps [54]. Hence, the behaviour of ΔE with respect to voltage change reveals the nature of the dominant traps involved in the conduction. An empirical model is used to fit the variation of the activation energy (ΔE) of traps with the bias voltage (V$_b$) to analyze the nature of traps in the BiFeO$_3$ films. Eq. (4) gives this model, and its response behaviour is shown in Fig. 7 (a).

$$\Delta E = (\Delta E_0 - \Delta E_1) + \Delta E_1 * exp(-V_b/V_d) \quad (4)$$

In the activation energy model, ΔE$_0$ serves as the zero bias activation energy, while ΔE$_1$ and V$_d$ represent the model's fitting parameters, with ΔE$_1$ being positive when characterizing donor-type traps and negative for acceptor-type traps. The ΔE extracted from Arrhenius curves at different bias voltages are shown in Fig. 7b for the respective BiFeO$_3$ films. The activation energy of BFO_700 films increases with bias, saturating to a value, while other devices exhibit an exponential decreasing trend. On fitting the curves in Fig. 7b with the empirical model (4), it is inferred that BFO_400, BFO_500 and BFO_600 predominantly feature donor-type traps, while BFO_700 indicates the presence of acceptor-type traps.

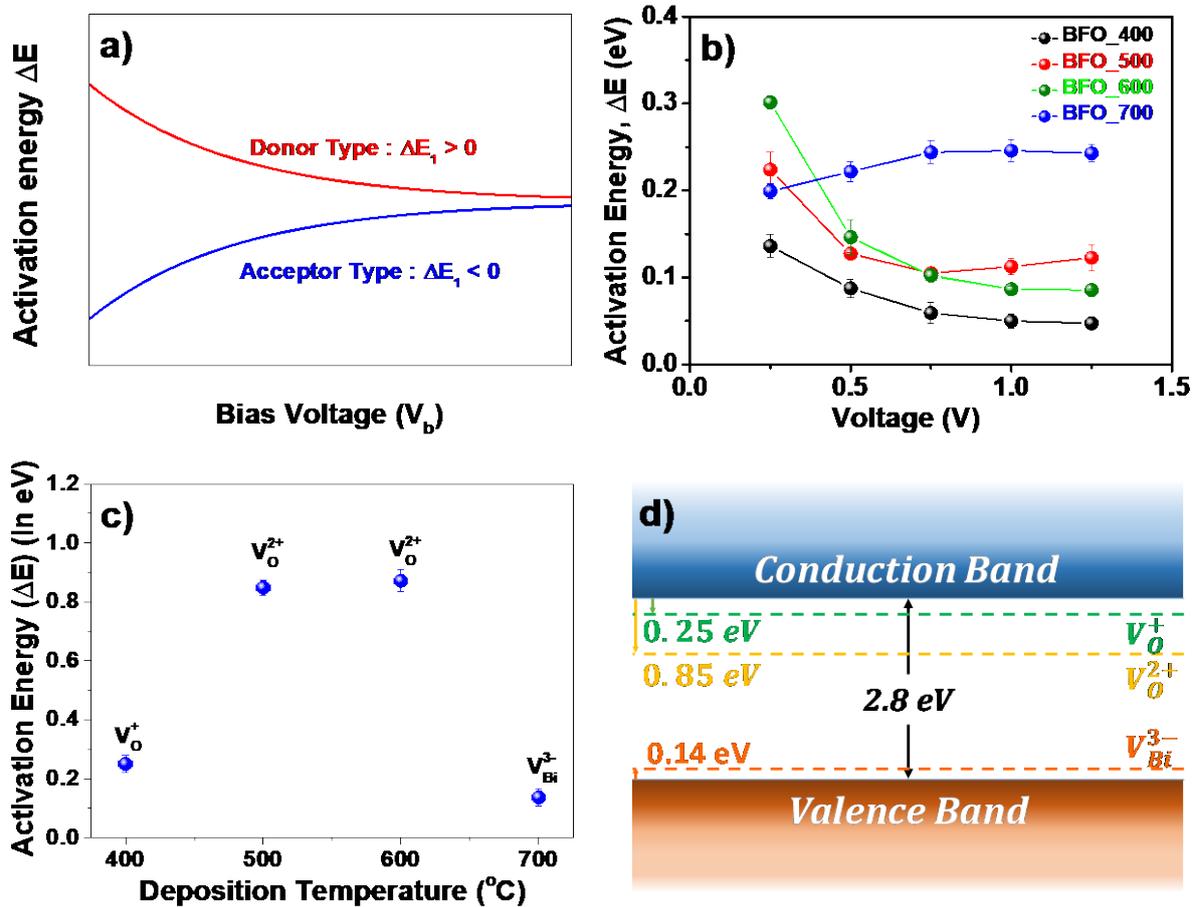

**Fig.-7:** (a): Empirical model of activation energy plotted as a function of the applied bias for acceptor and donor type defects (b): Activation energy of traps measured for BiFeO$_3$ film deposited at different temperatures, indicating films deposited at 700 °C having acceptor-type defect levels, (c): Zero bias activation energy of dominant traps at different deposition temperature and (d): Energy band diagram showing the relative location of defect levels

In the theoretical investigation of intrinsic point defects in BiFeO$_3$ by Shimada et al. [38], single and double valent oxygen vacancies ($V_O^+$ and $V_O^{2+}$) were identified as donor-type defects closer to the conduction band edge alongside bismuth vacancies ($V_{Bi}^{3-}$) as the acceptor-type defects near the valence band edge. The calculated zero-bias activation energy *(ΔE₀)* of associated traps in the BiFeO$_3$ films deposited at different temperatures are illustrated in Fig. 7c and the relative location of the traps in the energy band diagram is shown in Fig. 7d. The inference is that the predominant traps in 400°C films close to $V_O^{+}$ type, for 500°C and 600°C films, the activation energies match with that of $V_O^{2+}$ type defects. Hence, BFO_400, BFO_500 and BFO_600 films behave as an n-type device, where conduction is through oxygen vacancies. However, for 700°C devices, the activation energy confirms with $V_{Bi}^{3-}$ defects, making the film p-type with Bi vacancies [22]. This deduction is supported by the work of Rojac et al. [54], where it was reported that bismuth vacancies drive domain wall conduction in BiFeO$_3$ thin films. All the estimated key parameters from different experiments are summarized in Table 3.

| Estimated Parameters | BFO_400 | BFO_500 | BFO_600 | BFO_700 |
| --- | --- | --- | --- | --- |
| Oxygen Vacancies $V_o$ (%) | 14.58 | - | - | 2.81 |
| RMS Roughness $R_q$ (nm) | 3.24 | 5.6 | 8.09 | 15.11 |
| Optical Bandgap $E_g$ (eV) | 2.63 | 2.66 | 2.75 | 2.81 |
| Trap Density $N_T$ x $10^{16}$ (cm$^{-3}$) | 8.68 | 6.76 | 1.75 | 1.43 |
| Conduction Band Offset $\varphi_b$ (eV) | 0.69 | 0.76 | 0.78 | 0.96 |
| Nature of Trap | Donor | Donor | Donor | Acceptor |
| Trap Activation Energy $\Delta E$ (eV) | 0.25 | 0.85 | 0.87 | 0.14 |

**Table 3**: Consolidation overview of the key parameters in the study

## 4. Conclusion

In this work, we demonstrate the manipulation of defect type and density in BiFeO$_3$ thin film by varying the deposition temperature in PLD. The material analysis indicates the crystallinity and grain size improvement as the deposition temperature is higher and the oxygen content can be tuned by adjusting the deposition temperature. Current-voltage analysis of the $BiFeO_3$ film reveals that donor-type defects dominate at lower deposition temperatures, while acceptor-type defects are prevalent at higher temperatures. By controlling the carrier type and conductivity through deposition temperature modulation, $BiFeO_3$ becomes a promising material for a range of electronic applications, such as selector-free resistive memory devices and p-n junction neuromorphic devices. These findings provide a solid foundation for further research into the practical implementation of $BiFeO_3$-based devices.

## CRediT authorship contribution statement

**R S Viswajit:** Conceptualization, Formal Analysis, Investigation, Methodology, Visualization, Writing – original draft. **Kochupurackal B Jinesh:** Conceptualization, Methodology, Resources, Supervision, Validation, Writing-review & editing. **Ashok K:** Resources, Supervision, Writing-review & editing.

## Declaration of Competing Interest

The authors declare that they have no known competing financial interests or personal relationships that could have appeared to influence the work reported in this paper.

## Acknowledgement

We extend our gratitude to the teams who contributed to completing this work. Special thanks to our colleagues at the Emerald lab for their invaluable support in setting up and maintaining the PLD system. We thank the Department of Chemistry at IIST, Thiruvananthapuram, for supporting absorption and photoluminescence spectroscopy experiments. The Avionics departments at IIST, Thiruvananthapuram, helped us revalidate the

Absorption spectroscopy results. The Physics department at Kerala University, Kariyavattom, supported us in conducting XRD and XPS characterizations. Lastly, Dr. Vivekanand V, VSSC/ISRO for his review of the paper and valuable suggestions for organizing the content.

## Appendix A. Supplementary material

The detailed AFM results with 3D surface image, 2D image profile and line scan of BiFeO$_3$ films for 500°C and 600°C devices. Tau plots of the UV-VIS absorption spectrum of the thin films for optical bandgap estimation. Log I versus log V plots of all four films highlighting trap-filled limit voltage. The Schottky thermionic emission plots and Arrhenius plots at different voltage bias conditions for film realized at various deposition temperatures are included.